\documentclass[prd,twocolumn,reprint,preprintnumbers,nofootinbib,superscriptaddress]{revtex4-1}

\input{header}

\newcommand{\btau}{B-3\tau}
\newcommand{\bltau}{B-3L_\tau}

\begin{document}

\preprint{PITT-PACC-2122, UCI-TR-2021-16, DESY 21-191}

\title{
\vspace*{0.5in}
Hadrophilic Dark Sectors at the Forward Physics Facility
\vspace*{0.2in}
}

\author{Brian~Batell}
\affiliation{Pittsburgh Particle Physics, Astrophysics, and Cosmology Center, Department of Physics and Astronomy, University of Pittsburgh, Pittsburgh, PA 15217, USA}

\author{Jonathan L.~Feng}
\affiliation{Department of Physics and Astronomy,  University of California, Irvine, CA 92697-4575, USA}

\author{Max Fieg}
\affiliation{Department of Physics and Astronomy,  University of California, Irvine, CA 92697-4575, USA}

\author{Ahmed Ismail}
\affiliation{Department of Physics, Oklahoma State University, Stillwater, OK, 74078, USA}

\author{Felix Kling}
\affiliation{Theory Group, SLAC National Accelerator Laboratory, Menlo Park, CA 94025, USA}
\affiliation{Deutsches Elektronen-Synchrotron DESY, Notkestrasse 85, 22607 Hamburg, Germany}

\author{Roshan Mammen Abraham}
\affiliation{Department of Physics, Oklahoma State University, Stillwater, OK, 74078, USA}

\author{Sebastian Trojanowski
\vspace*{0.2in}
}
\email{batell@pitt.edu, jlf@uci.edu, mfieg@uci.edu, \\ aismail3@okstate.edu, felix.kling@desy.de,\\  rmammen@okstate.edu,  strojanowski@camk.edu.pl}
\affiliation{Astrocent, Nicolaus Copernicus Astronomical Center Polish Academy of Sciences, ul.~Rektorska 4, 00-614, Warsaw, Poland}
\affiliation{National Centre for Nuclear Research, ul.~Pasteura 7, 02-093 Warsaw, Poland
\vspace*{0.25in}
}

\begin{abstract}
Models with light dark sector and dark matter particles motivate qualitatively new collider searches.  Here we carry out a comprehensive study of hadrophilic models with U(1)$_B$ and U(1)$_{\bltau}$ gauge bosons coupled to light dark matter.  The new mediator particles in these models couple to quarks, but have suppressed couplings to leptons, providing a useful foil to the well-studied dark photon models.  We consider current bounds from accelerator and collider searches, rare anomaly-induced decays, neutrino non-standard interactions, and dark matter direct detection.  Despite the many existing constraints, these models predict a range of new signatures that can be seen in current and near future experiments, including dark gauge boson decays to the hadronic final states $\pi^+ \pi^- \pi^0$, $\pi^0 \gamma$, $K^+ K^-$, and $K_S K_L$ in FASER at LHC Run 3, enhancements of $\nu_{\tau}$ scattering rates in far-forward neutrino detectors, and thermal dark matter scattering in FLArE in the HL-LHC era.  These models therefore motivate an array of different experiments in the far-forward region at the LHC, as could be accommodated in the proposed Forward Physics Facility.
\end{abstract}

\maketitle

\section{Introduction}
\label{sec:intro}

Searches for new particles and dark matter (DM) are primary physics drivers at the Large Hadron Collider (LHC).  Traditional searches for the classic missing $p_T$ signature at the LHC main detectors have sensitively searched for particles with weak-scale masses and ${\cal O}(1)$ couplings to Standard Model (SM) particles, but are less effective for light and weakly coupled new particles, including long-lived particles (LLPs) and DM.  Recently it has been appreciated that new experiments in the far-forward region at the LHC can provide a powerful probe of new light particles.  These experiments exploit the large forward flux of pions and other SM particles, which, if they decay to new light particles, can create a large forward flux of LLPs and DM. Light new physics species can also be produced in the far-forward region of the LHC in other types of interactions, including proton-proton bremsstrahlung and the Drell-Yan process. The recent detection of TeV neutrino candidates in the forward region~\cite{FASER:2021mtu} also opens a new window on neutrinos at colliders, which may be used to probe both SM and beyond the SM (BSM) phenomena~\cite{Abreu:2019yak, Ahdida:2020evc, Anchordoqui:2021ghd}.

In evaluating any proposal for new physics at the MeV to GeV mass scale, one must carefully consider all of the existing constraints from particle and nuclear experiments carried out over the last 60 years.  To do this requires a model framework.  The dark photon model has been discussed at length in the literature.  It is theoretically attractive and contains within it phenomenologically-viable benchmark scenarios of light thermal DM.  Of particular relevance for this study, previous studies in the dark photon framework have established the potential for forward experiments to detect both LLPs~\cite{Feng:2017uoz,Ariga:2018uku} and light thermal DM~\cite{Batell:2021blf,Batell:2021aja}.  At the same time, the experimental signatures of a given dark sector model are, to a large extent, determined by the interactions of the mediator with the SM. To more fully evaluate the physics potential of proposed experiments, then, a variety of phenomenologically distinct mediators must be examined. Since the LHC is a $pp$ collider, it is natural to consider mediators with hadrophilic couplings, i.e., sizable couplings to quarks, but suppressed couplings to leptons. Although such models are challenging to test at electron facilities (e.g., Belle-II~\cite{Belle-II:2018jsg}, NA64~\cite{Banerjee:2019dyo}, LDMX~\cite{Akesson:2018vlm}, and SENSEI~\cite{Tiffenberg:2017aac}), one might suspect that they can be sensitively probed at proton facilities, such as the LHC.  

In this work we study the prospects for probing two dark sector models with hadrophilic vector boson mediators. The first model is based on a gauged U(1)$_B$ baryon number symmetry~(see, e.g., Refs.~\cite{Dobrescu:2014fca, Tulin:2014tya, Batell:2014yra, Soper:2014ska, Dobrescu:2014ita}). This model is perhaps the first example of a hadrophilic model one might consider, since it has sizable couplings to quarks and (loop-)suppressed couplings to all leptons.  The model suffers from gauge anomalies, however, which potentially lead to stringent constraints from rare FCNC and $Z$ boson decays~\cite{Dror:2017ehi,Dror:2017nsg}. We will evaluate the prospects for discovering new physics in this model, carefully respecting all anomaly constraints, as well as those from other experimental searches. We note that anomaly-free extensions of the SM with a local U(1)$_B$ symmetry and DM have been constructed in Refs.~\cite{FileviezPerez:2010gw,Duerr:2013dza,Duerr:2013lka,FileviezPerez:2014lnj}, which focus on the case of new particle masses above the weak scale.

As a second example we consider a model with a U(1)$_{\bltau}$ vector boson mediator.  (In the rest of this paper, we will use the modest abbreviation of U(1)$_{\btau}$ for this symmetry.) With the addition of a right-handed neutrino, this symmetry is anomaly free and therefore evades the most stringent rare decay constraints present in the U(1)$_B$ model. This model is also hadrophilic, in the sense that couplings to electrons, muons, and their accompanying neutrinos are suppressed. However, the presence of $\tau$ and $\nu_{\tau}$ couplings brings with it both additional constraints from neutrino nonstandard interactions (NSI), and also new opportunities for signals involving the 3rd generation leptons. A goal of this study is to incorporate all these new constraints and see what discovery prospects remain.

We will consider both current and proposed far-forward experiments.  In the last two years, the magnetic spectrometer and tracking detector FASER~\cite{Ariga:2018pin}, and the two emulsion detectors FASER$\nu$~\cite{Abreu:2020ddv} and SND@LHC~\cite{Ahdida:2750060} have been approved.  FASER has been fully constructed, and all three are expected to begin taking data when Run 3 starts in 2022.  For the High Luminosity LHC (HL-LHC) era, detectors under consideration include upgrades of these detectors (FASER2, FASER$\nu$2, and Advanced SND), as well as the Forward Liquid Argon Experiment (FLArE)~\cite{Batell:2021blf}.\footnote{As a potential upgrade of milliQan~\cite{Haas:2014dda}, a fifth experiment, FORMOSA~\cite{Foroughi-Abari:2020qar}, has also been proposed to carry out dedicated searches for milli-charged particles and similar signatures.}  A new facility, the Forward Physics Facility (FPF)~\cite{SnowmassFPF,Anchordoqui:2021ghd}, has been proposed to accommodate these experiments.

Remarkably, we will find that all of these detectors have discovery prospects for the hadrophilic models we consider.  The possible signals include DM deep inelastic scattering (DIS) and elastic scattering, enhanced predictions for neutrino neutral current (NC) scattering, an excess of tau neutrinos in the forward region, and the visible decay of the dark mediators into SM final states.  Notably, the visible decays include final states, such as $\pi^+ \pi^- \pi^0$, $\pi^0 \gamma$, $K^+ K^-$, and $K_S K_L$, that could conceivably appear in FASER at LHC Run 3; such states are inaccessible at FASER in dark photon models.  The signals are diverse and require a similarly diverse set of experiments to find them, and when combined, the experiments probe parameter space even beyond the DM thermal targets.  These models therefore add to the broad physics portfolio of the FPF, complementing other studies of long-lived particle searches, collider-produced TeV-energy neutrinos, new probes of QCD, and high-energy astroparticle physics~\cite{Anchordoqui:2021ghd}. 

The paper is organized as follows. In \secref{model} we introduce the two hadrophilic dark sector models based on the U(1)$_B$ and U(1)$_{\btau}$ gauge symmetries and discuss the production and decays of the vector boson mediator, the DM thermal relic abundance, and the existing constraints for each model. Next, we present our assumptions regarding the performance of FASER, FASER2, SND@LHC, FASER$\nu$2, and FLArE in \secref{detectors}. In \secref{signatures} we outline our methodology for estimating the sensitivity of these far-forward detectors to the new physics signatures predicted in these hadrophilic models.  Our main results are contained in \secref{result}, and our conclusions and outlook are presented in \secref{conclusions}.

\section{Models of Hadrophilic Physics}
\label{sec:model}

\subsection{Models}

With the motivation outlined in \secref{intro}, we begin in this section by describing the two representative hadrophilic dark sector models based on the anomalous U(1)$_B$ and anomaly-free U(1)$_{\btau}$ gauge symmetries.\footnote{The cancellation of gauge anomalies in the U(1)$_{\btau}$ model requires the introduction of a right-handed neutrino with ${\btau}$ charge of $-3$. In this study we assume that the right handed neutrino is somewhat heavier than the vector boson mediator, which can be achieved by coupling it to the dark Higgs field that spontaneously breaks U(1)$_{\btau}$. In principle the heavy neutrino mass could reside anywhere in the range below $m_V/g_V$. Depending on its mass and mixing with SM neutrinos there could be additional signatures beyond the core phenomenology outlined below.  These are beyond the scope of our study, but see Ref.~\cite{Jodlowski:2020vhr} for the sub-$\gev$ case and far-forward searches.}  
Since the new gauge group is Abelian, the new vector gauge boson generically mixes with the SM photon through a kinetic mixing term $F_{\mu\nu} V^{\mu\nu}$, where $F_{\mu\nu}$ and $V_{\mu\nu}$ are the field strengths of the SM photon and new gauge boson, respectively.  In the physical mass basis, the Lagrangian of the vector boson mediator $V_\mu$ is 
\begin{equation}
\label{eq:L-V}
{\cal L} \supset -\frac{1}{4} V_{\mu\nu} V^{\mu\nu}+\frac{1}{2} m_V^2 V_\mu V^\mu + V_\mu (J^\mu_{\text{SM}} + J^\mu_\chi) \ ,
\end{equation}
where $m_V$ is the vector boson mass, $J^\mu_{\text{SM}}$ is a current composed of SM fields, and $J^\mu_\chi$ is the current for the dark matter particle $\chi$.

The SM current is
\begin{equation}
J^{\mu}_{\text{SM}} = 
g_V [J^\mu_B - 3 x (\overline \tau \gamma^\mu \tau +  \overline \nu_\tau \gamma^\mu P_L \nu_\tau) ] + \varepsilon \, e \, J^\mu_{\text{EM}} \ ,
\end{equation}
where $g_V \equiv \sqrt{4 \pi \alpha_V}$ is the new U(1) gauge coupling, $J_B^\mu$ and $J^\mu_{\text{EM}}$ are the baryon number and electromagnetic currents, respectively, $\varepsilon$ is the kinetic mixing parameter, and $x= 0$ (1) for the U(1)$_B$ (U(1)$_{\btau}$) model. 

To specify $J^\mu_\chi$, we must choose the DM candidate $\chi$. We will study both complex scalar DM and Majorana fermion DM in this work, with Lagrangians
\be
\!\!\!\mathcal{L} \supset 
\begin{cases}
\displaystyle{\, |\partial_\mu \chi|^2 \!-\! m_\chi^2 |\chi|^2} \, ,
\quad \quad \ \ \text{complex scalar} \\
\vspace{-10pt}\\
\displaystyle{\, \frac{1}{2} \overline \chi i \gamma^\mu \partial_\mu \chi 
\!-\!\frac{1}{2} m_\chi \overline \chi \chi} \, , 
\ \, \text{Majorana fermion} \ ,  \\
\end{cases}
\!\!\!\!
\ee
where $m_\chi$ is the DM mass. The associated currents, $J^\mu_\chi$ in \cref{eq:L-V}, are
\be
\label{eq:JD}
J_\chi^\mu = g_V Q_\chi
\begin{cases}
\displaystyle{\ i \chi^* \overset{\text{\footnotesize$\leftrightarrow$}}{\partial^\mu} \chi} \, , \quad \ \text{complex scalar}  \\
\vspace{-10pt}\\
\displaystyle{\ \frac{1}{2} \overline \chi \gamma^\mu \gamma^5 \chi}  \, ,
\ \text{Majorana fermion}  \ , \\
\end{cases}
\ee
where $Q_\chi$ is the charge of the DM under the new gauge symmetry. As we will discuss below, both complex scalar and Majorana fermion DM exhibit velocity-suppressed $P$-wave annihilation to SM final states, implying that bounds from precision measurements of the cosmic microwave background anisotropies~\cite{Ade:2015xua, Slatyer:2009yq} are easily satisfied in these models. Furthermore, Majorana DM features momentum-dependent scattering in the non-relativistic regime, making it challenging to probe with DM direct detection experiments. This is not the case for complex scalar DM, and, as we will see, direct detection experiments place strong constraints on such DM for masses above the GeV scale. However, it is important to note that these constraints can also be evaded in a straightforward way by introducing a small mass splitting, which renders the scattering transition inelastic~\cite{Han:1997wn,Hall:1997ah,TuckerSmith:2001hy}.  

The full parameter space of these models is, then, specified by 5 parameters:
\begin{equation}
m_V , \ g_V , \ \varepsilon , \ m_\chi , \text{ and } Q_\chi \ .
\end{equation}
To reduce the parameter space, as is commonly done in the literature, we will assume a kinetic mixing parameter of typical one-loop size,
\begin{equation}
\varepsilon = \frac{e \, g_V}{16 \pi^2} \ .
\label{eq:epsilonrelation}
\end{equation}
This is the parametric size of the kinetic mixing generated by loops of SM particles charged under both electromagnetism and the new gauge symmetry. The kinetic mixing depends, in general, on the details of the UV physics and therefore cannot be determined unambiguously, but we neglect such effects here; see also Ref.~\cite{Bauer:2018onh} for further discussion of this issue. Throughout our study we will also adopt another common convention,
\begin{equation}
    m_V = 3 m_\chi \ ,
    \label{eq:massrelation}
\end{equation}
so that DM annihilation proceeds to SM particles through a virtual $s$-channel vector boson mediator.  

Given the assumptions of \eqsref{epsilonrelation}{massrelation}, the resulting parameter space may be specified by the three parameters
\begin{equation}
m_V , \ g_V , \text{ and } Q_\chi \ .
\end{equation}
We will present our results in the $(m_V, g_V)$ plane with various choices for $Q_{\chi}$.  Since the new symmetries are Abelian, the charge $Q_\chi$ may be any real number. When presenting our results below, we will consider two choices for coupling hierarchies. As a first scenario, we will consider DM and SM particles to have comparable interaction strengths with the vector boson mediator, fixing
\begin{equation}
\label{eq:Qchi}
Q_\chi =  
\begin{cases}
1 \, , \  \text{U(1)$_B$ models} \\
3 \, , \  \text{U(1)$_{\btau}$ models} \ . 
\end{cases}
\end{equation}
In the $\btau$ model, we have fixed the DM charge to be opposite that of the $\nu_\tau$, $Q_\chi = -Q_\tau$. As a second, qualitatively distinct, scenario, we consider the case in which the DM coupling to the vector boson mediator has a fixed value, 
\begin{equation}
\alpha_\chi \equiv \frac{g_V^2 Q_\chi^2}{4 \pi} = 0.01 \text{ or } 0.5 \ . 
\label{eq:largeQ}
\end{equation}
Given that we will be considering vector boson mediators with weak couplings to the SM, that is, values of $g_V \sim 10^{-8} - 10^{-2}$, \eqref{largeQ} implies very large DM charges $Q_{\chi}$.  This may appear unnatural, but there is nothing wrong in principle, since the expansion parameter $\alpha_\chi$ remains perturbative.  Ideas for achieving such large coupling hierarchies for two U(1) gauge symmetries have been presented in Ref.~\cite{Berlin:2018bsc}.

Finally, although we do not consider them in this work, viable models of hadrophilic scalar mediators can also be constructed; see, e.g., Refs.~\cite{Batell:2017kty, Batell:2018fqo, Batell:2021xsi}.  However, for the incident DM energies in the TeV range relevant for FPF experiments, scalar-mediated DM-nuclear scattering rates are typically suppressed by several orders of magnitude in comparison to vector boson-mediated scattering rates; see also Ref.~\cite{Berlin:2018bsc} for a comparison of vector boson- and scalar-mediated DM scattering in the ultra-relativistic regime. For this reason, scalar-mediated DM scattering can be better probed by low- and medium-energy experiments~\cite{Batell:2018fqo}. On the other hand, experiments such as FASER and FASER2 can have powerful sensitivity to visible decays of the long-lived scalar mediator in these models, as has been demonstrated in Ref.~\cite{Kling:2021fwx}. 

\subsection{Production and Decay of the Vector Boson Mediator\label{sec:proddecay}}

In our simulations, we model the production of light dark vector bosons in the far-forward region of the LHC by employing the \texttt{FORESEE} package~\cite{Kling:2021fwx}. We thereby include dark vector boson production by light meson decays, proton bremsstrahlung,\footnote{The modeling of dark vector boson production via proton bremsstrahlung in \texttt{FORESEE} is based on the Fermi-Weizsacker-Williams approximation presented in Refs.~\cite{Blumlein:2013cua,deNiverville:2016rqh}. Recently, Ref.~\cite{Foroughi-Abari:2021zbm} studied this process using an alternative model of nucleon interactions based on Pomeron exchange, finding production rates that are smaller by a factor of a few. These estimates provide a sense of the theoretical uncertainty inherent in this process.
}
and the Drell-Yan process. We observe that typically the production of dark vector bosons in light meson decays dominates if kinematically allowed. For the dark vector bosons heavier than the $\eta$ meson, the most important production mode is due to bremsstrahlung, while the Drell-Yan process starts to dominate for $m_V > 1.5~\gev$.

We then consider various decay final states of the dark vector bosons. In particular, the partial decay width for $V \to \chi \chi^*$ is 
\begin{equation}
\Gamma_{\chi \chi^*} =\kappa \, \frac{\alpha_\chi  m_V}{12}\left(1-\frac{4 m_\chi^2}{m_V^2}\right)^{3/2},
\label{eq:Vdecay}
\end{equation}
where $\kappa$ = 1 and 2 for complex scalar and Majorana DM, respectively. The partial decay width into hadrons and other SM particles is taken from the \texttt{DarkCast} package~\cite{Ilten:2018crw}, which used data-driven methods to estimate the hadronic width. An alternative description has also recently been implemented in \texttt{Herwig~7}; see Ref.~\cite{Plehn:2019jeo}.

\begin{figure*}
\centering
\includegraphics[width=0.49\textwidth]{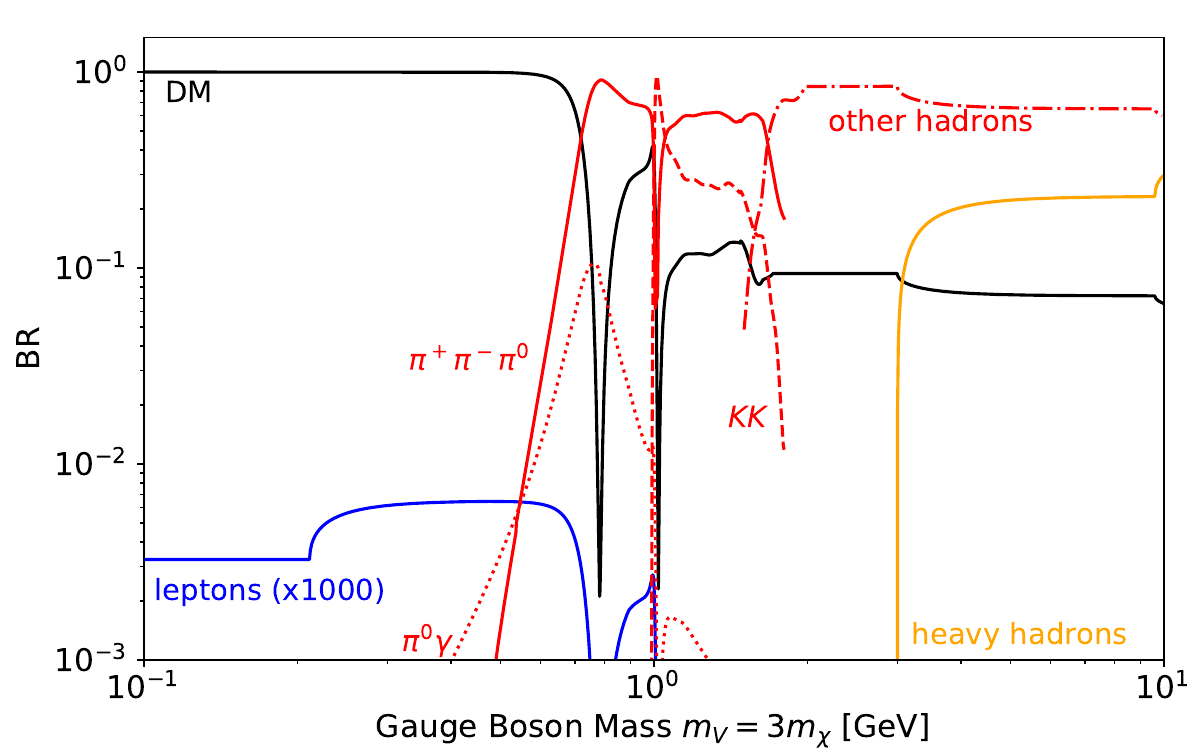}
\includegraphics[width=0.49\textwidth]{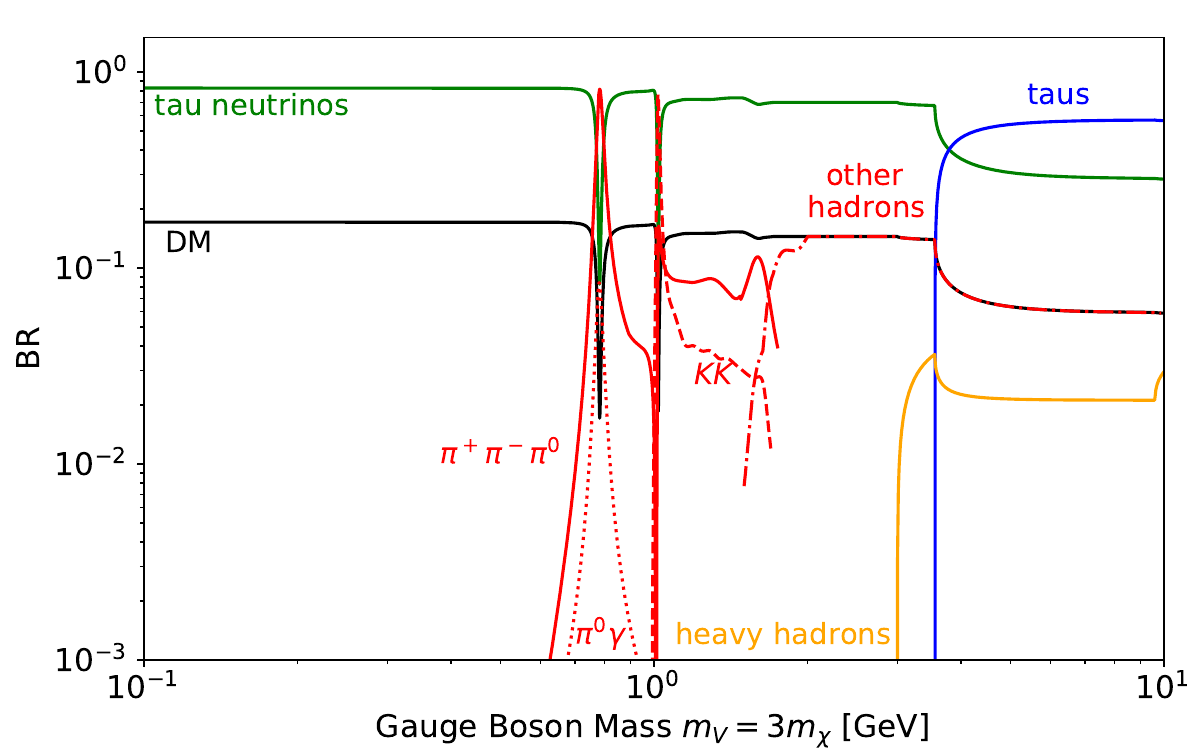}
\caption{Decay branching fractions of the $B$ (left) and $\btau$ (right) gauge bosons for fixed $Q_\chi = 1$ and $3$, respectively. The ``heavy hadrons'' contour includes charm and bottom hadrons, and the red contours correspond to all other hadrons. Among them, we explicitly show the dominant branching fractions into $\pi^0\pi^+\pi^-$, $\pi^0\gamma$, and kaon pairs $KK = K^+ K^- + K_S K_L$. Here we assume loop-induced couplings of the bosons to charged leptons of the first two generations of size $g_\ell = g_V \, (e/4\pi)^2$. The relevant contour for boson decays into $e^+ e^-$ or $\mu^+ \mu^-$, shown in the left panel, has been multiplied by a factor of $1000$ for visibility. The DM is taken to be a scalar, with the decay width given in \cref{eq:Vdecay}.
\label{fig:BR} }
\end{figure*}

In \cref{fig:BR}, we present the corresponding decay branching fractions for both of the models assuming the $Q_\chi$ charge as in \cref{eq:Qchi}. 
In the case of the U(1)$_B$ model, LLP decays into lepton pairs are always subdominant, since they appear only at the loop level through the vector boson mixing with the photon. In contrast, the invisible branching fraction of $V\to \chi\chi^*$ is close to unity for light vector boson masses up to the $\omega$-resonance region, $m_V \approx m_\omega \simeq 782~\mev$. This leads to an intense flux of DM particles, which can be detected via DM scatterings, as we will discuss in \cref{sec:DIS,sec:elastic}. For heavier dark vector bosons, decays into light hadrons start to play an important role and can lead to additional signatures in the detectors, as we will see in \cref{sec:decaysignature}.  

For the $\btau$ model with the dark charge set to $Q_\chi = 3$, we obtain $\textrm{BR}(V\to\chi\chi^*) \sim (10-20)\%$ up to the tau threshold, above which $V \to \tau^+ \tau^-$ decays become kinematically allowed. The remaining decay rate for lighter dark vector bosons is dominantly into tau neutrinos, $V \to \nu_\tau \overline {\nu}_\tau$. As will be discussed in \cref{sec:tausignature}, this can contribute to the total $\nu_\tau$ flux measured at the FPF. The decays into hadrons also become important for certain values of $m_V$, especially around the $\omega$- and $\phi$-resonance regions.

\subsection{Thermal Relic Abundance}

Thermal targets, that is, the regions of parameter space where DM annihilates in the early Universe through thermal freezeout to the correct relic density, provide an important standard by which to judge the sensitivity of collider searches.  These have been determined in the U(1)$_B$ model with fixed $\alpha_{\chi} = 0.5$ in Ref.~\cite{Berlin:2018bsc}.  Here we determine, for the first time, the thermal targets for the U(1)$_B$ model with fixed $Q_{\chi}$ and for the U(1)$_{\btau}$ model described above.  

The dark matter annihilation cross section can be written in the standard resonance form, 
\begin{equation}
\sigma_{\rm ann} (s)= \kappa\,\frac{16 \pi}{s \beta_\chi^2} \frac{(2 s_V +1)}{(2s_\chi+1)^2 }\frac{s \ \Gamma_{\chi \chi^*}(s) \ \Gamma_{\rm SM}(s)}{(s-m_V^2)^2 + m_V^2 \Gamma_V^2} \, ,
\label{eq:sigma_ann}
\end{equation}
where $\beta_\chi(s) = (1-4 m_\chi^2 /s)^{1/2}$, $s_V = 1$, $s_\chi = 0$, and  $\Gamma_{\chi \chi^*}(s)$ and $\Gamma_{\rm SM}(s)$ are the partial decay widths for $V$ decaying into dark matter and SM particles, respectively, with the replacement $m_V \rightarrow \sqrt{s}$. 

The thermally-averaged cross section is, then,~\cite{Gondolo:1990dk}
\begin{equation}
    \langle \sigma_{\rm{ann}} v\rangle = \frac{\kappa}{2} \frac{\int_{4m_{\chi}^2}^{\infty} \sqrt{s} (s \! - \! 4m_{\chi}^2) \, \sigma_{\rm{ann}}(s) \, K_1(\sqrt{s}/T) \, ds}{8m_{\chi}^4T K^2_2(m_{\chi}/T)} \, ,
    \label{eq:sigma_ann_averaged}
\end{equation}
where $v$ is the relative velocity of the annihilating dark matter particles, and $K_i$ is the modified Bessel function of order $i$. To determine the thermal target regions of parameter space, we require
\begin{equation}
\langle \sigma_{\rm ann} v\rangle  = 4 \times 10^{-26}~\cm^3~\text{s}^{-1} \ ,
\end{equation}
which reproduces the observed DM relic abundance for the masses we consider~\cite{Steigman:2012nb}. 

The thermal targets are presented below in \cref{fig:gequal,fig:fixedalphaD}.   Their shapes can be understood as follows.  In the U(1)$_{\btau}$ models, annihilation to tau neutrinos is allowed throughout the $m_V$ range.  The thermally-averaged cross section has the parametric dependence
\begin{equation}
\label{eq:sigv-parametric}
\langle \sigma_{\text{ann}} v \rangle 
\sim \frac{\kappa g_V^4 Q_\chi^2}{ m_V^2} 
\sim \frac{\kappa g_V^2 \alpha_\chi}{m_V^2}  \ ,   
\end{equation}
and so in the $(\log m_V, \log g_V)$ plane, the thermal targets have slope 1 for the models with fixed $\alpha_\chi$ shown in \figref{fixedalphaD}, and slope 1/2 for the models with fixed $Q_\chi$ shown in \figref{gequal}.  The discrepancy between the complex scalar and Majorana fermion cases results from the fact that in the complex scalar case, there are both DM and anti-DM particles, whereas in the Majorana case, DM is its own anti-particle, which impacts the annihilation rate through the parameter $\kappa$'s appearance in \eqsref{sigma_ann}{sigma_ann_averaged}.

For the U(1)$_B$ models, the thermal target slopes are similar to those for the U(1)$_{\btau}$ models for $m_V \agt 1~\gev$. The required couplings $g_V$ are greater because the annihilation to tau neutrinos is absent.  As $m_V$ drops below 1 GeV, the cross section to hadrons decreases rapidly, and without a large leptonic annihilation channel, the required $g_V$ increases rapidly to maintain a fixed $\langle \sigma_{\text{ann}} v \rangle$.  This continues until $m_V$ drops below $m_{\pi}$, at which point all hadronic channels shut off, and only the loop-suppressed annihilation to light leptons is allowed. The curve moves further up for masses $m_V/3 = m_\chi < m_e$ where only the high velocity tail of the thermal DM population can annihilate into electrons, which needs to be compensated by a larger coupling. However, even though only a small fraction of DM can annihilate into electrons, this is still more efficient than the annihilation into 3 photons. The latter process, $\chi\chi \to 3 \gamma$~\cite{Pospelov:2008jk,McDermott:2017qcg}, was found to be negligible for our study.

The resonance structure seen in all cases arises from resonant mixing of the dark gauge boson $V$ with the SM vector mesons $\rho$, $\omega$, and $\phi$. In the case of DM annihilation, these resonances occur at masses $2 m_{V}/3 = 2 m_{\chi} = m_{\rho,\phi,\omega}$, whereas for $V$ production, these resonances occur at $m_V = m_{\rho,\phi,\omega}$.

\subsection{Existing Constraints}
\label{sec:constraints}

Light hadrophilic mediators have a rich phenomenology, giving rise to constraints from previous searches, as well as search opportunities at FPF experiments. Below, we summarize the various laboratory experimental constraints on light hadrophilic gauge bosons following the discussion of Ref.~\cite{Kling:2020iar}. The resulting limits are shown in \cref{fig:fixedalphaD,fig:gequal} as dark gray shaded regions.

\begin{description}

\item [Invisible Mediator Decays] The focus of this study are hadrophilic mediators with a sizable branching fraction into dark matter. This decay leads to missing energy signatures which have been searched for by various experiments. The most sensitive constraints have been obtained by the search for the decay $\pi^0 \to \gamma V$ at NA62~\cite{CortinaGil:2019nuo} and LESB~\cite{Atiya:1992sm}; the search for the decay $\pi^0,\eta,\eta' \to \gamma V$ at Crystal Barrel~\cite{Amsler:1994gt}; the search for the decay $K^+ \to \pi^+ V$ at E949~\cite{Artamonov:2009sz} as discussed in Ref.~\cite{Pospelov:2008zw, Batell:2014yra}; the search for the mixing induced invisible decays of the $J/\Psi$ by BES~\cite{BES:2007sxr} and the $\Upsilon$ by BaBar~\cite{BaBar:2009gco} as discussed in Ref.~\cite{Batell:2014yra}; and the monojet search $pp \to V+\text{jet}$ at CDF~\cite{Aaltonen:2012jb} as discussed in Ref.~\cite{Shoemaker:2011vi}. 
 
\item [Visible Mediator Decays] If the couplings of the hadrophilic mediator to the SM and dark sector have similar size, decays into visible final states are possible. If the coupling is sufficiently large, the decays of the mediator occur promptly in the detector and can be searched for via a bump hunt. Bounds have been obtained by the search for the decay $\eta' \to  V \gamma \to \pi^0 \gamma \gamma $ at GAMS-2000~\cite{GAMS:1987kiw} and the search for the non-electromagnetic contribution to the decay $\Upsilon(1S) \to jj$ by ARGUS~\cite{ARGUS:1986nzm}, as discussed in Ref.~\cite{Aranda:1998fr}. In addition, there are bounds from searches for displaced decays of LLPs from NuCAL~\cite{Blumlein:1990ay}. 
 
\item [DM and Neutrino Scattering] The hadrophilic mediator is copiously produced in beam dump experiments. The decay $V \!\to\! \chi\chi^*$ then leads to a dark matter beam. The MiniBooNE collaboration has searched for the scattering of $\chi$ in their downstream neutrino detector~\cite{MiniBooNE:2017nqe,MiniBooNEDM:2018cxm}. Recently, even stronger bounds on coherent scatterings of leptophobic DM have been obtained with the Coherent CAPTAIN-Mills (CCM) liquid argon (LAr) detector~\cite{Aguilar-Arevalo:2021sbh}. Similarly, the decay $V \!\to\! \nu_\tau \overline \nu_\tau$ leads to an increased tau neutrino flux, which can be constrained using measurements from DONuT~\cite{Kodama:2007aa}, as discussed in Ref.~\cite{Kling:2020iar}. 
 
\item [Indirect Probes] A hadrophilic mediator can also be constrained indirectly through its contribution to the low-energy neutron-lead scattering cross section~\cite{Barbieri:1975xy}, as discussed in Ref.~\cite{Barger:2010aj}.
Additionally, a new gauge boson with couplings to tau leptons can be constrained by the measurement of the $Z \to \tau\tau$ decay width at LEP~\cite{Tanabashi:2018oca}, as discussed in Ref.~\cite{Ma:1998dp}.

\end{description}

In addition, there are other constraints that are somewhat more model dependent.  These are the anomaly constraints and the constraints from neutrino NSIs, which are shown as light gray shaded regions in \figsref{fixedalphaD}{gequal}, and which we now describe:

\begin{description}
 \item [Anomaly Constraints] As mentioned above, the dark vector boson in the U(1)$_B$ model couples to a non-conserved SM current. Invisible decays of such a vector boson are then constrained by enhanced bounds from missing energy searches in rare $Z$ decays and flavor-changing meson decays $K \to \pi V$ and $B \to K V$. We implement them following Refs.~\cite{Dror:2017ehi,Dror:2017nsg}, assuming that anomalies associated with the new gauge group are canceled by heavy fermions that do not receive masses from electroweak symmetry breaking. If these anomalies were canceled by fermions with Yukawa couplings to the Higgs, the invisible decay constraints would not apply, but there would be severe LHC constraints on the additional fermions.

 \item [Neutrino NSI] For the U(1)$_{\btau}$ model, additional constraints arise from studying neutrino oscillations, both in vacuum and in the matter background of the Sun and Earth. These have been precisely measured by a variety of neutrino experiments. A global fit to these neutrino oscillations measurements simultaneously constrains the oscillation parameters and NSI between neutrinos and matter. We present these bounds following Ref.~\cite{Han:2019zkz}.\footnote{An alternative study, which obtained slightly stronger constraints, was performed in Ref.~\cite{Heeck:2018nzc} using the global fit results obtained in Ref.~\cite{Esteban:2018ppq}.} We note, however, that these constraints are model dependent and could be weakened in the presence of additional new physics.

 \item [Direct Detection] Further bounds on hadrophilic DM can arise from direct detection (DD) searches~\cite{Batell:2014yra}. These, however, depend sensitively on the detailed structure of the DM interaction and do not apply to Majorana DM and to inelastic scalar DM if the mass gap between the dark species is large enough to suppress upscatterings of non-relativistic DM particles. We stress this in the following when presenting the current DD bounds on spin-independent DM-nuclei scattering from the CRESST-III~\cite{CRESST:2019jnq}, DarkSide-50~\cite{DarkSide:2018bpj}, and Xenon 1T~\cite{XENON:2018voc, XENON:2020gfr} experiments. We show these bounds assuming that $\Omega_\chi h^2\simeq 0.12$~\cite{Planck:2018vyg} in the entire reach plot and that a non-standard cosmological scenario affects the DM relic density for points in the parameter space away from the thermal target lines.
 
 \item [Cosmology \& Astrophysics] Further indirect probes arise from possible contributions of light dark vector bosons to the number of relativistic degrees of freedom in the early Universe, $\Delta N_{\textrm{eff}}$. We present them below following Refs.~\cite{Escudero:2019gzq, Bauer:2020itv}. Additional bounds could arise from an enhanced supernova cooling rate of SN1987A, as discussed, for example, in Refs.~\cite{Dent:2012mx, Dreiner:2013mua, Kazanas:2014mca, Rrapaj:2015wgs, Chang:2016ntp, Chang:2018rso, Sung:2019xie}.  Such constraints typically probe very small couplings outside the regions of interest for this study. In addition, they are also dependent on a number of astrophysical assumptions, which may weaken the constraints or possibly even remove them altogether; see, e.g., Ref.~\cite{Bar:2019ifz}. In the following, we do not show these bounds explicitly in our sensitivity reach plots, as a detailed study for the models considered here is beyond the scope of our analysis.

\end{description}

\section{Detectors \label{sec:detectors}}

We perform our analysis for the on-axis far-forward detectors that will operate either during LHC Run 3 or the HL-LHC era. In the latter case, we focus on the proposed FPF, which begins at a distance $L= 620~\m$ away from the ATLAS Interaction Point~\cite{Anchordoqui:2021ghd}. In particular, we study the expected future sensitivity of the $10$-tonne emulsion detector FASER$\nu$2, a proposed successor to the FASER$\nu$ experiment that will take data during LHC Run 3~\cite{Abreu:2019yak, Abreu:2020ddv}, as well as the $10$- and $100$-tonne fiducial mass liquid-argon time projection chamber (LArTPC) detectors FLArE-10 and FLArE-100~\cite{Batell:2021blf}. The relevant detector geometries are
\be
\text{FASER$\nu$2:} \ \ &\Delta = 2~\m,  &S_T& = (0.5~\m \times 0.5~\m)  , 
\\
\text{FLArE-10:} \ \    &\Delta = 7~\m,  &S_T& = (1~\m \times 1~\m)  ,
\\
\text{FLArE-100:} \ \   &\Delta = 30~\m, &S_T& = (1.6~\m \times 1.6~\m) ,
\label{eq:detector1}  
\nonumber
\ee
where $\Delta$ is the length of the detector, and $S_T$ denotes its transverse size.

Both types of detectors have excellent capabilities to reconstruct the low-energy nuclear scattering signals created by both neutrinos and hadrophilic DM, and also to disentangle DM-induced events from the more energetic neutrino scatterings. For these searches, however, it is also important that they be able to reject backgrounds induced by high-energy muons that pass through the facility and interact with the surrounding rock and infrastructure. To veto these muons, it is highly beneficial to collect time information about the events. In the case of FASER$\nu$2, this would likely require interleaving the emulsion layers with additional electronic detectors. For FLArE, on the other hand, the required time resolution can be more easily obtained by employing an additional light collection system; see Ref.~\cite{Batell:2021blf} for further discussion. 

Throughout this paper, we use the neutrino fluxes for the FPF as presented in Ref.~\cite{Batell:2021aja}. These fluxes were obtained using the event generator \texttt{SIBYLL~2.3d}~\cite{Ahn:2009wx, Riehn:2015oba, Riehn:2017mfm, Riehn:2019jet} as implemented in \texttt{CRMC}~\cite{CRMC} to simulate the primary collision, and the fast neutrino flux simulation presented in Ref.~\cite{Kling:2021gos} to model the propagation and decay of long-lived SM hadrons in the forward LHC infrastructure.

In addition to the aforementioned scattering detectors, we will also present sensitivities for the LLP signature of the vector boson mediator decaying to visible SM final states. To this end, we will focus on FASER~\cite{Ariga:2018zuc, Ariga:2018pin} and FASER2, cylindrical detectors with length $\Delta$ and radius $R$, where~\cite{Ariga:2018uku}
\be
\text{FASER:} \ \   &\Delta = 1.5~\m, &R& = 10~\cm, &\mathcal{L}& = 150~\ifb\, ,
\\
\text{FASER2:} \ \ &\Delta = 5~\m,   &R& = 1~\m,   &\mathcal{L}& = 3~\iab\, .
\label{eq:detector2}\nonumber
\ee
FASER will take data during LHC Run 3 and will be positioned in the far-forward region at a distance $L=480~\m$ away from the ATLAS IP.  For FASER2 we assume the relevant parameters for the HL-LHC era and the FPF location. Above, we have also provided the relevant integrated luminosities. The multiple collisions that occur in each bunch crossing (pile-up) are accounted for in determining the flux of $V$.

Throughout the study, we assume perfect detection efficiency for all the events that pass the selection criteria. The probability of passing such criteria depends on the geometrical acceptance of the detectors, energy and other kinematic cuts, as well as on the final state interactions inside the nucleus that we take into account in the case of the elastic scattering. We discuss the relevant cuts for different signatures below.

We will also include in our plots the expected sensitivities of the SND@LHC detector~\cite{SHiP:2020sos} to DM scattering in the U(1)$_B$ model, as determined in Ref.~\cite{Boyarsky:2021moj}. For the elastic DM scattering signature, this analysis assumed that backgrounds from muon-induced hadrons and photons can be rejected and that the number of neutrino-induced events can also be suppressed to a negligible level.  In the DIS regime, the analysis estimated that pure neutrino-induced backgrounds could be reduced to $\mathcal{O}(1000)$ events, and the sensitivity curves were taken to be $N = 100$ DM signal event contours. 

\section{Signatures \label{sec:signatures}}

\renewcommand{\arraystretch}{1.2}
\begin{table*}[tbp]
\setlength{\tabcolsep}{5.2pt}
\centering
\begin{tabular}{c|c|c|c|c|c}
  \hline \hline
  \textbf{Signature}  
  & \textbf{DM DIS}
  & \textbf{DM Elastic} 
  & \textbf{$\nu$ NC DIS} 
  & \textbf{$\nu_\tau$ CC DIS}    
  & \textbf{LLP decays} 
  \\ 
  \hline
  \textbf{Section}  
  & \cref{sec:DIS}
  & \cref{sec:elastic}
  & \cref{sec:BSMneutrinos}
  & \cref{sec:tausignature}    
  & \cref{sec:decaysignature} 
  \\ 
  \hline
  \textbf{Models}      
  & $U(1)_B$,  $U(1)_{B-3\tau}$ 
  & $U(1)_B$, $U(1)_{B-3\tau}$  
  & \hspace{0.43cm} $U(1)_{B-3\tau}$ \hspace{0.43cm}          
  & \hspace{0.43cm} $U(1)_{B-3\tau}$ \hspace{0.43cm}
  & $U(1)_B$, $U(1)_{B-3\tau}$ 
  \\ 
  \hline
  \textbf{Production}
  & $pp \to V \to \chi\chi$
  & $pp \to V \to \chi\chi$
  & $pp \to D_s \to \nu_\tau$
  & $pp \to V \to \nu_\tau \bar{\nu}_\tau$
  & $pp \to V$
  \\ 
  & \includegraphics[width=0.15\textwidth]{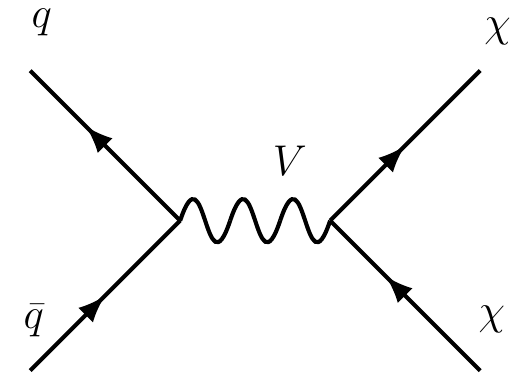} 
  & \includegraphics[width=0.15\textwidth]{DM_production.pdf} 
  & \includegraphics[width=0.15\textwidth]{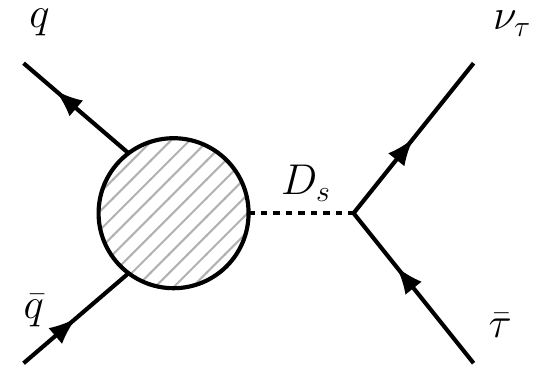} 
  & \includegraphics[width=0.15\textwidth]{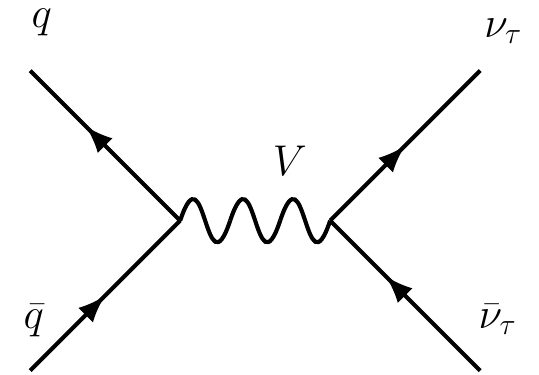} 
  & \includegraphics[width=0.092\textwidth]{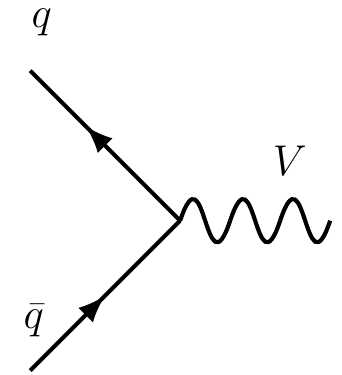}
  \\ 
  \hline
  \textbf{Detection}
  & $\chi  N \to \chi X$
  & $\chi  p \to \chi p$
  & $\nu_{\tau}  N \to \nu_{\tau}  X$
  & $\nu_{\tau}  N \to \tau  X$
  & $V \to\ $hadrons
  \\ 
  & \includegraphics[width=0.1\textwidth]{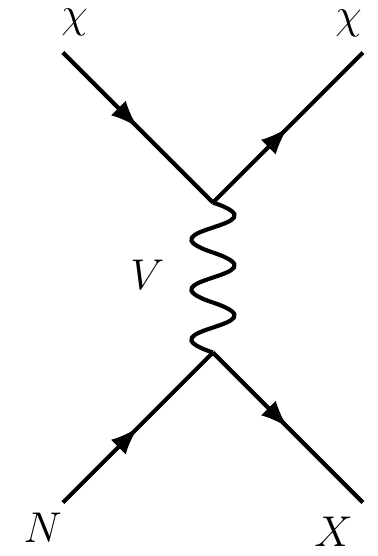} 
  & \includegraphics[width=0.1\textwidth]{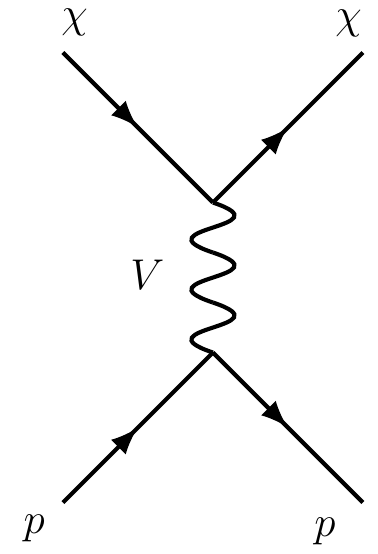} 
  & \includegraphics[width=0.1\textwidth]{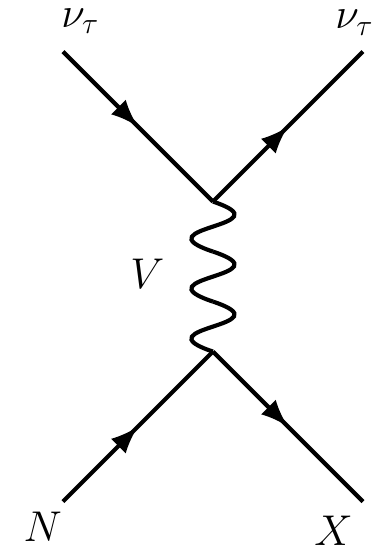} 
  & \includegraphics[width=0.1\textwidth]{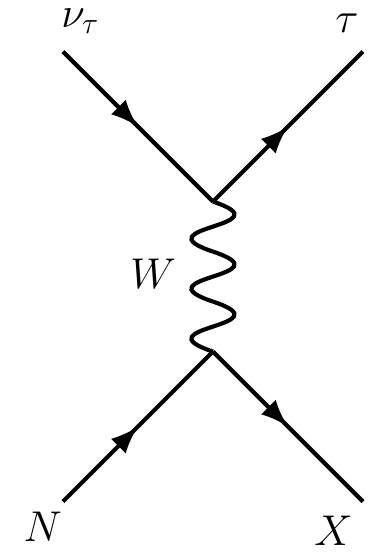} 
  & \begin{overpic}[scale=0.30, tics=5, trim=0 -1.8cm -1cm 0, clip]{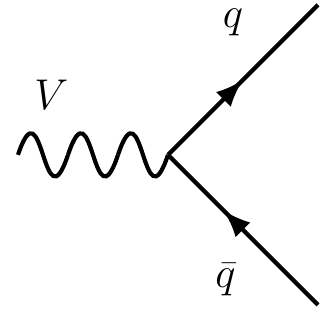}
  \end{overpic}
  \\
  \hline
  \textbf{Rate scales as}
  & $ g_V^6 Q_\chi^2 \sim g_V^4 \alpha_\chi$ 
  & $ g_V^6 Q_\chi^2 \sim g_V^4 \alpha_\chi$ 
  & $ g_V^4$ 
  & $ g_V^2$
  & \!\!\!\!$ g_V^2 e^{-g_V^2 m_V^2}$ or $g_V^4$\!\!\!\!
  \\
  \hline
  \textbf{Background}    
  & $\nu \ N \rightarrow \nu \ X$
  & $\nu \ p \rightarrow \nu \ p$
  & $\nu \ N \rightarrow \nu \ X$
  & $D_s \rightarrow \nu_{\tau} \tau$
  & None                                                          \\ 
  \hline
  \hline
\end{tabular}

\caption{The signatures studied. In the first three rows, the name of the signature, the subsection in which it is discussed, and the relevant new physics models are given. In the 4th and 5th rows, we show the Feynman diagrams for some example production and detection processes, respectively.  The production processes shown are not necessarily the dominant ones. The 6th row shows the dependence of the signal rate on the model parameters, and the 7th row lists the dominant SM backgrounds.}
\label{tab:summary}
\end{table*}

The hadrophilic models we are considering produce a diverse array of new physics signatures. These are shown in \cref{tab:summary}, where we list which models are relevant for each signature, the dominant production and detection processes that determine the signal rates, the dependence of these rates on the model parameters, and the dominant SM backgrounds. As can be seen, the FPF experiments will be sensitive to direct signals generated by both the dark vector boson and DM, as well as to neutrino-induced signals. We now discuss them all in detail. 

\subsection{DM Deep Inelastic Scattering\label{sec:DIS}}

We first consider DM DIS off nuclei, $\chi N \to \chi X$. At large momentum transfer, DM DIS produces a significant hadronic recoil with multiple charged tracks. The main background is SM neutrino NC interactions. Due to the light mediator, DM scattering prefers lower momentum transfer than the neutrino background, which proceeds through $Z$-boson exchange. Our discussion of this signature closely follows that in Ref.~\cite{Batell:2021aja}. 

The differential cross section for complex scalar DM DIS in the models of Sec.~\ref{sec:model} is given by 
\be
\label{eq:disxsec}
&&\frac{d\sigma(\chi N \!\to\! \chi X)}{dx \ dy} 
= 4 \pi \alpha_\chi \alpha_V  \frac{2 m_p E_\chi}{(Q^2+m_{A'}^2)^2}\ \times \\
&&\ \sum_{q=u,d,s,c}\!\! (1 - y) \big[ x f_{q}(x,Q^2) + x f_{\bar{q}}(x,Q^2) \big ] \, , 
\ee
where $x$ is the parton momentum fraction, $y = 1 - E'_{\chi} / E_{\chi}$ is the fraction of the incoming DM energy transferred to the nucleon in the lab frame, $Q^2 = 2 m_p E_\chi x y$ is the squared momentum transfer, and $f_q$ is the quark parton distribution function. We use the nCTEQ15 parton distribution functions~\cite{Kovarik:2015cma} for tungsten and argon and integrate \cref{eq:disxsec} requiring $Q^2 > 1~\gev^2$ to obtain the expected numbers of DM DIS events in the FPF detectors. We also require the energy transferred to the hadronic system to be $1~\gev \!<\! E_\text{had} \!<\! 15~\gev$, where $E_\text{had} = y E_\chi$, and the total transverse momentum of the recoiling hadrons to be $1~\gev \!<\! p_{T,\text{had}} \!<\! 1.5~\gev$, where $p_{T,\text{had}}^2 = Q^2 (1 \!-\! y)$. For the background, we calculate the expected numbers of neutrino NC scattering events satisfying the same cuts on $Q^2$, $E_\text{had}$, and $p_{T,\text{had}}^2$. Our cuts favor softer hadronic recoils, eliminating much of the neutrino NC background. Our projected sensitivities assume perfect detector efficiency and consider only statistical uncertainties. A previous study~\cite{Batell:2021aja} of DM DIS at FLArE found that some experimentally motivated cuts did not have a large effect on the signal, but a full study remains to be performed.

\subsection{DM-Nucleon Elastic Scattering\label{sec:elastic}}

The light DM particles produced in the far-forward region at the LHC can also be discovered via their elastic scatterings with nucleons, which lead to single proton tracks visible in the detector. We treat this signature following Ref.~\cite{Batell:2021aja}, in which we have also studied the relevant neutrino-induced backgrounds. In particular, when presenting the sensitivity contours, we require the momentum of the outgoing proton to be within the range $300~\mev < p_p < 1~\gev$ in FASER$\nu$2, and for FLArE we require $p_p < 1~\gev$ and the proton's kinetic energy to satisfy $E_{k,p}>20~\mev$. We also reject events in which other visible tracks emerge from the vertex. After these cuts, we expect $\sim 100$, 1000, and 300 background events during the entire HL-LHC run for FLArE-10, FLArE-100, and FASER$\nu$2, respectively.

The elastic scattering cross section for the complex scalar DM interacting with the neutron or proton via the hadrophilic gauge boson is 
\be
\label{eq:dffxsecDMQE-U1B}
& \frac{d\sigma(\chi p \to \chi p)}{dQ^2} = \frac{4 \pi \alpha_\chi \alpha_V Q^2}{(E_\chi^2 - m_\chi^2)(m_{V}^2+Q^2)^2} \times \!\Big\{ A(Q^2)  \\
& \quad\quad 
 + \left(\frac{E_\chi}{Q} \!-\! \frac{Q}{4m_N}\right)^2 
\!\!\left[(\tilde{F}^B_{1,N})^2 \!+\! \tau (\tilde{F}^B_{2,N})^2\right] \!\Big\},
\ee
where $Q^2 = 2\,m_N\,(E_N\!-\!m_N)$ is the squared four-momentum transfer in terms of the nucleon mass $m_N$ ($N=n,p$) and the outgoing nucleon energy $E_N$, and $E_\chi$ corresponds to the incident DM energy. The term proportional to $A(Q^2)$, which contributes negligibly to the cross section at high energies, is given by
\begin{equation}
A(Q^2) =
-\frac{1}{4}\,(\tilde{F}^B_{1,N}+\tilde{F}^B_{2,N})^2\,\left(\tau+\frac{m_\chi^2}{m_p^2}\right),
\label{eq:AfactorU1B}
\end{equation}
with $\tau = Q^2/(4 m_p^2)$. In contrast to the case of a vanilla dark photon mediator, the neutron and proton form factors are identical in this case and given by
\begin{align}
\tilde F^B_{1,N}(Q^2) & = \frac{ 1 + (\mu_p+\mu_n)\,\tau}{1+\tau}\, G_D(Q^2)\ , \\
\tilde F^B_{2,N}(Q^2) & = \frac{\mu_p+\mu_n -1 }{1+ \tau }\, G_D(Q^2) \ ,
\end{align}
where $\mu_p = 2.793$, $\mu_n = -1.913$, and $G_D(Q^2) = (1+ Q^2/M^2)^{-2}$, with $M = 0.843\, {\rm GeV}$. The differential elastic scattering cross section becomes form-factor suppressed at large momentum transfers, and the total elastic cross section is dominated by the contribution from $Q^2\lesssim m_{V}^2$.

In the following, we include scatterings off both protons and neutrons. For protons, we include the efficiency factors $\sim(50-70)\%$ related to the final-state interactions (FSI) of protons, as in Ref.~\cite{Batell:2021aja}. For neutrons, we include similar efficiency factors in the range $(15-30)\%$, which have been obtained as a function of the outgoing neutron momentum by studying neutrino interactions in \texttt{GENIE}~\cite{Andreopoulos:2009rq,Andreopoulos:2015wxa}. In this case, the neutron re-scatterings inside the nucleus can lead to an outgoing proton with momentum within the aforementioned cuts and with no other detectable tracks. We find that scatterings of DM off neutrons can contribute up to 25\% to the total elastic event rate.

\subsection{Enhanced Neutrino Neutral Current Scattering \label{sec:BSMneutrinos}}

When a new mediator couples to neutrinos, NC scattering $\nu N \to \nu X$ receives an additional contribution from the mediator. The signature is identical to that for DM DIS. However, as NC scattering depends only on the couplings of the mediator to quarks and neutrinos, there is no dependence on $m_\chi$ or $Q_\chi$, unlike the case of DM scattering. In particular, for the $\btau$ mediator, the total $\nu_\tau$ NC cross section becomes 
\be
\label{eq:nuncxsec}
&\frac{d\sigma(\nu N \!\to\! \nu X)}{dx \ dy} 
= \frac{m_p E_{\nu}}{4 \pi} \times  \\
& \sum_{q=u,d,s,c} \! \! \! \left\{ c^{2}_{L} \, \big[ x f_{q}(x,Q^2) + x(1 - y)^2 f_{\bar{q}}(x,Q^2) \big ] \right. \\
& \left. \quad \quad \ \  + c^{2}_{R} \, \big[ x(1 - y)^2 f_{q}(x,Q^2) + x f_{\bar{q}}(x,Q^2) \big ] \right\} ,
\ee
where
\be
c_{L/R} &= \frac{(g_W \, g_{\nu ,L})(g_W \, g_{q ,L/R})}{\cos^2 \! \theta_W (Q^2 + m^{2}_Z)} + \frac{1}{4}\frac{(g^{2}_{V} Q_{\nu} Q_{q})}{(Q^2 + m^{2}_V)} \ .
\ee
Here $g_W$ is the SM weak coupling, $g_{\nu,L} = \frac{1}{2}$, and $g_{q,L}= \frac{1}{2} - \frac{2}{3} \sin^2 \! \theta_W$ for up-type quarks and $-\frac{1}{2} + \frac{1}{3} \sin^2 \! \theta_W$ for down-type quarks. The second term in $c_{L,R}$ is the contribution from the new $\btau$ mediator with charges $Q_\nu = -3$ (3) for $\nu_\tau$ ($\bar{\nu}_\tau$), and $Q_q = \frac{1}{3}$ for all quarks. The interference term is proportional to $Q_{\nu}Q_{q}$, and so carries opposite signs for $\nu_\tau$ and $\overline{\nu}_{\tau}$ NC scattering~\cite{Dev:2021xzd}. At the FPF where we expect almost equal fluxes of $\nu_\tau$ and $\bar{\nu}_{\tau}$, this implies a small contribution from the interference term after cancellations. Nevertheless we use the complete expression above in our analysis.

For small $m_V$, the BSM contribution to NC scattering prefers low recoil energy, similar to DM DIS and unlike the weak boson-mediated SM process, whose cross section grows with momentum transfer. We calculate the number of additional NC events expected at the FPF with \cref{eq:nuncxsec}, using the same parton distribution functions and minimum $Q^2$ cut as in \cref{sec:DIS}. Because of the small relative flux of tau neutrinos compared to muon and electron neutrinos, the impact of the light mediator on the total NC cross section must be significant to provide a sizable effect relative to the SM NC background.

In testing whether an excess of NC events is observable, we consider only statistical uncertainties and neglect systematic uncertainties. For simplicity, we also assume perfect detection efficiency for NC interactions; the inclusion of realistic detection efficiencies~\cite{Ismail:2020yqc} would not substantially change the positions of the limits from excess NC events in \cref{fig:fixedalphaD,fig:gequal}, relative to the other signatures that we consider. We note that the main systematic uncertainty in the NC cross section measurement, the neutrino flux, can be constrained by measurements of charged current (CC) interactions. We find a statistically significant effect from the BSM contribution to NC scattering when the coupling to mass ratio of the new interaction is comparable to that of the weak interaction, $g_V / m_V \gtrsim g_W / m_W \approx 10^{-2}~\gev^{-1}$.

\subsection{Excess of Tau Neutrino Flux\label{sec:tausignature}}

In the case of the gauged $\btau$ scenario, the hadrophilic mediator decays into tau neutrinos with a sizable branching fraction.\footnote{Additional $\nu_\tau$ flux can be produced via $V$ decays into tau leptons for $m_{V}\gtrsim 2\,m_{\tau}$. However, the corresponding expected sensitivity lies in a region of parameter space that is already excluded, as shown in \cref{sec:result}.} As discussed in Ref.~\cite{Kling:2020iar}, this opens another opportunity to probe this model via their contribution to the LHC tau neutrino flux. In the SM, tau neutrinos are mainly produced via $D_s \to \nu \tau$ and subsequent $\tau$ decays, which occurs in roughly one in $10^5$ collisions at the LHC. This means that even rare BSM processes could lead to sizable contributions to the tau neutrino flux. The relevant detection channel in this case is via $\nu_\tau$ CC scatterings off nuclei. The displaced decays of the outgoing boosted tau lepton must then be identified in the detector, requiring excellent spatial resolution. 

An important issue that arises when searching for signs of new physics is the large uncertainty on the normalization of the SM tau neutrino flux~\cite{Bai:2020ukz, Kling:2021gos}. Although future efforts are expected to reduce these uncertainties, we will follow a different approach. In contrast to tau neutrinos from charm and tau decays, which have a broader angular spread, tau neutrinos from light mediator decays are more centered around the beam collision axis. In this study, we use this feature and perform a shape analysis of the $\nu_\tau$ angular distribution, which does not rely on knowledge of the neutrino flux normalization. We focus on the FLArE-10 design, whose $1~\m \times 1~\m$ cross sectional area is sufficiently large to capture this effect. More precisely, we define five concentric rectangular bins centered around the beam collision axis and corresponding to the distance between $d$ and $d+10~\cm$ away from it, where $d=0, 10, 20, 30, 40~\cm$. In practice, the most important contribution to the BSM-induced excess of $\nu_\tau$s is from the two most central bins, i.e., at distances up to $d\lesssim 20~\cm$ away from the beam collision axis.

\subsection{Visible Decays of the Dark Vector Boson\label{sec:decaysignature}}

In the following, we study the decay signature using the \texttt{FORESEE} package~\cite{Kling:2021fwx} with the lifetimes modeled with \texttt{DarkCast}~\cite{Ilten:2018crw} and the spectrum of light hadrons obtained from EPOS-LHC~\cite{Pierog:2013ria}. We assume $100\%$ detection efficiency for all visible final states. We present the results for both FASER and FASER2. In the analysis, we require the total energy of the visible products of the vector boson decays to be at least $100~\gev$. This cut has a minor impact on the BSM signal events, but suppresses possible SM backgrounds to a negligible level~\cite{Ariga:2018zuc,Ariga:2018pin}. Visibly decaying dark vector bosons could also appear in secondary production processes due to DM scatterings occurring right in front of or even inside the detector~\cite{Jodlowski:2019ycu}. We neglect the impact of such processes below, as we do not expect them to improve the sensitivity reach of the FPF detectors in the models under study.

\section{Results}
\label{sec:result}

\begin{figure*}[t]
\centering
\includegraphics[width=0.49\textwidth]{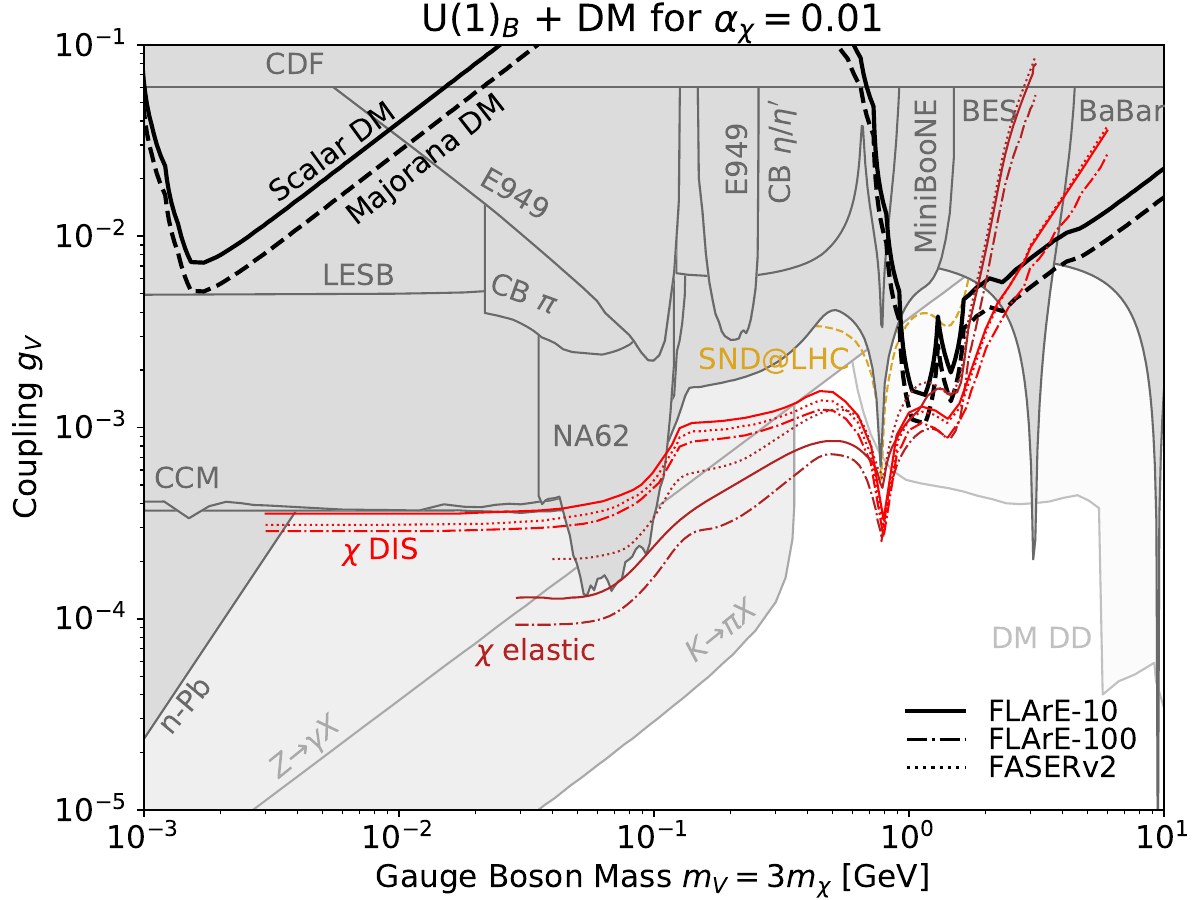}
\includegraphics[width=0.49\textwidth]{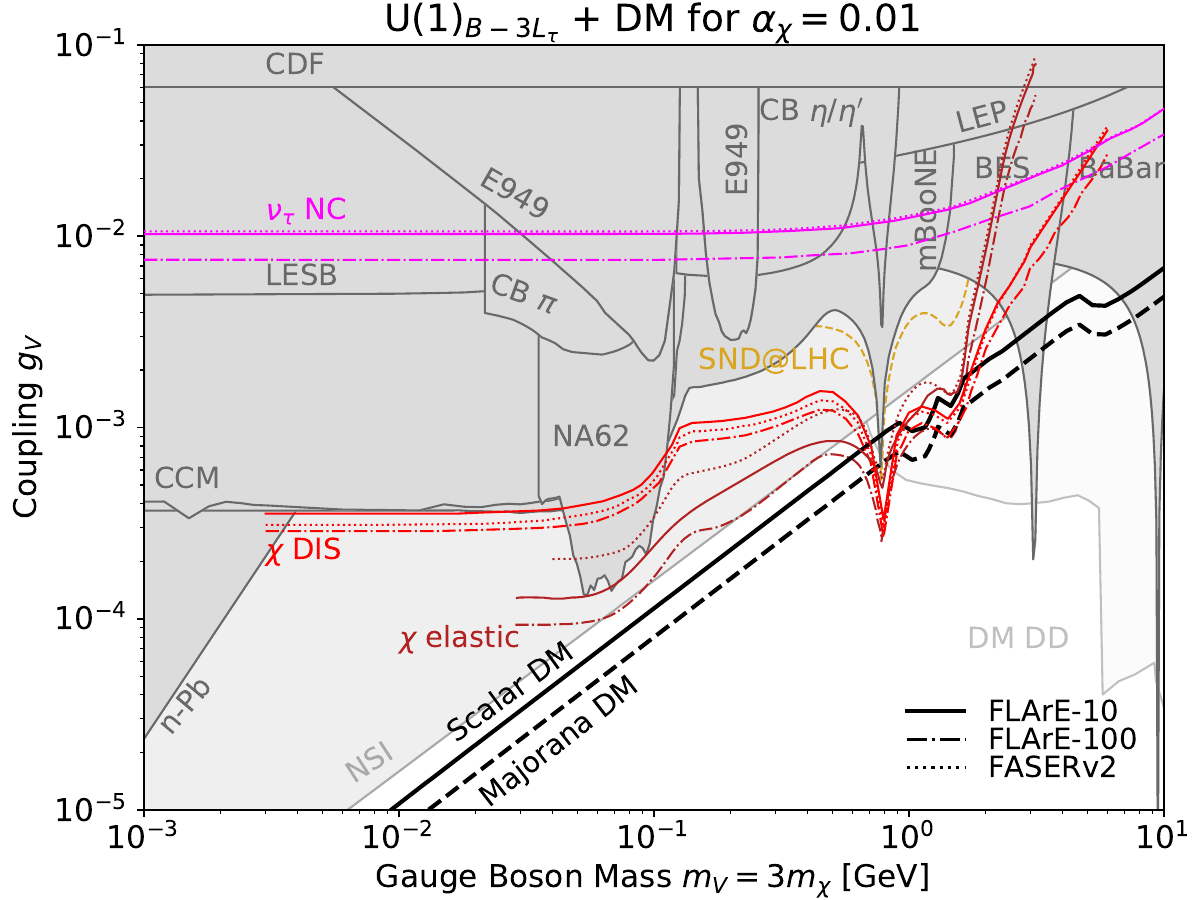}
\includegraphics[width=0.49\textwidth]{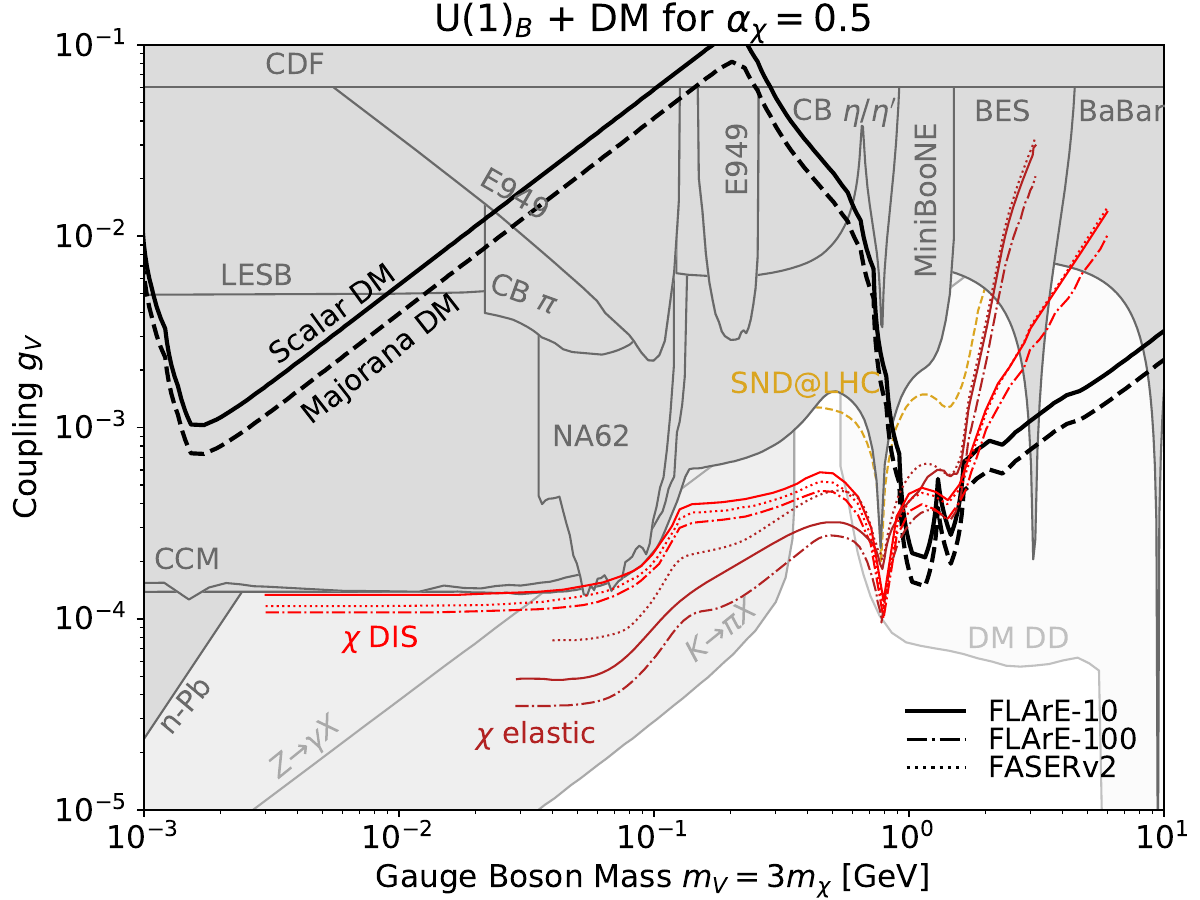}
\includegraphics[width=0.49\textwidth]{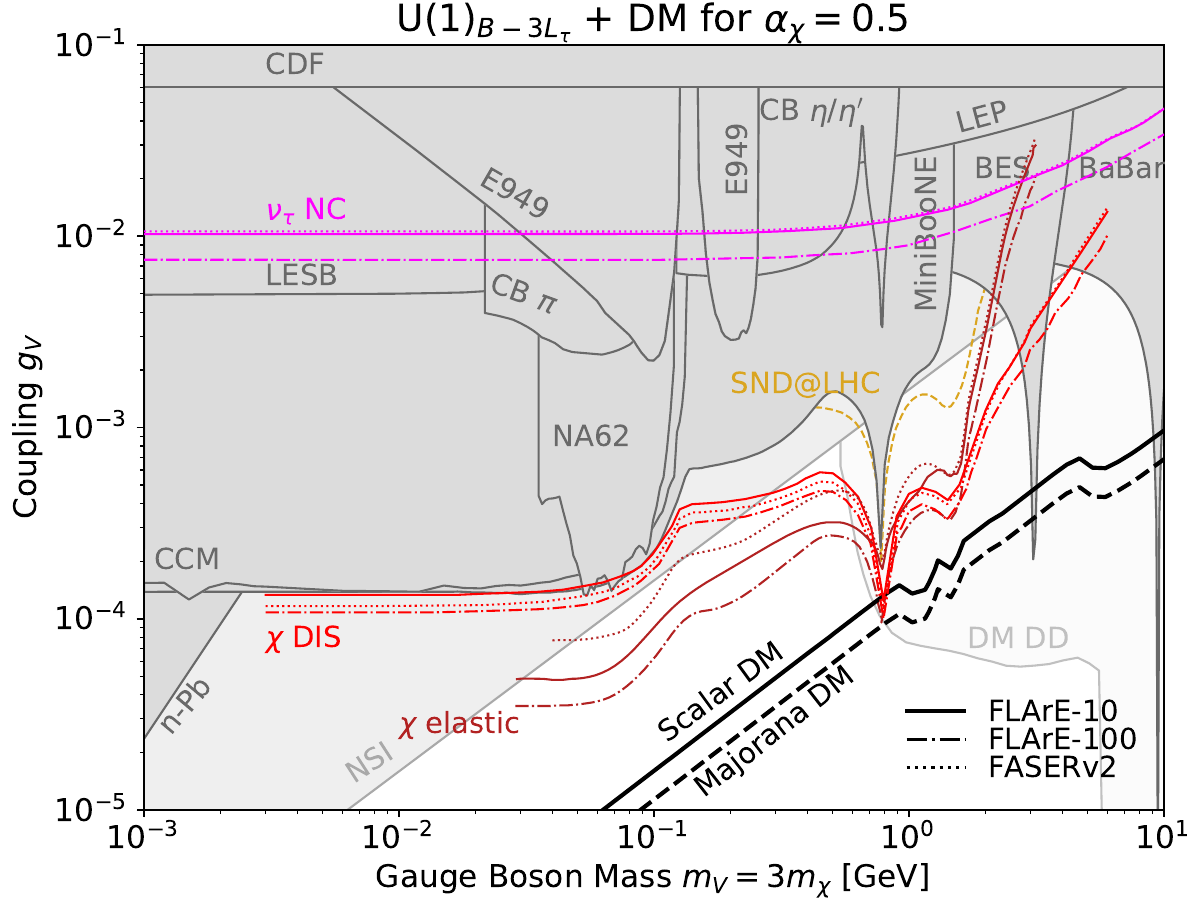}
\caption{The $(m_V,g_V)$ parameter space of hadrophilic DM models with U(1)$_B$ (left) and U(1)$_{\btau}$ (right) gauge boson mediators coupling to complex scalar DM, for dark matter coupling $\alpha_\chi = 0.01$ (top) and 0.5 (bottom), and $m_V = 3 m_\chi$.  The black contours are the thermal relic targets for complex scalar and Majorana DM; DM is thermally overproduced below these contours. The light (dark) red lines correspond to 90\% CL exclusion bounds from DM DIS (elastic) scatterings off nuclei for FLArE-10, FLArE-100, and FASER$\nu$2, as indicated. The dotted brown contours are the sensitivity contours for SND@LHC~\cite{Boyarsky:2021moj}. In the right panels, the light purple contours are the projected sensitivity contours from probing the $V$-induced BSM NC interactions of tau neutrinos. In both panels, the dark gray shaded regions are excluded by current bounds. The light gray shaded regions in the left (right) panels correspond to the anomaly-induced $K$ and $Z$ decays (NSI bounds).  The very light gray shaded regions are constraints from DM DD; these do not apply to Majorana and inelastic scalar DM (see \cref{sec:constraints}). } 
\label{fig:fixedalphaD}
\end{figure*}

In \cref{fig:fixedalphaD}, we present the results of our analysis for both the U(1)$_B$ and U(1)$_{\btau}$ models in the $(m_V,g_V)$ plane. In the plots, we fix the DM coupling to $\alpha_\chi = 0.01$ and $0.5$ in the upper and lower panels, respectively, and we keep a constant mass ratio between the dark sector particles, $m_V = 3\,m_\chi$. In dark gray, we show the existing constraints, as discussed in \cref{sec:constraints}, while the black solid (dashed) lines correspond to the relic density targets for the complex scalar (Majorana) DM. We stress that, although the anomaly bounds, shown in light gray in the left panels for the U(1)$_B$ case, can be avoided in modified versions of this simplified scenario, this often leads to further constraints due to additional couplings of the dark vector bosons that are introduced in the model to make it anomaly-free. An example is shown in the right panels for the anomaly-free U(1)$_{\btau}$ model, where the NSI constraints cover a good portion of the parameter space shown in the plot. 

\begin{figure*}[t]
    \centering
    \includegraphics[width=0.49\textwidth]{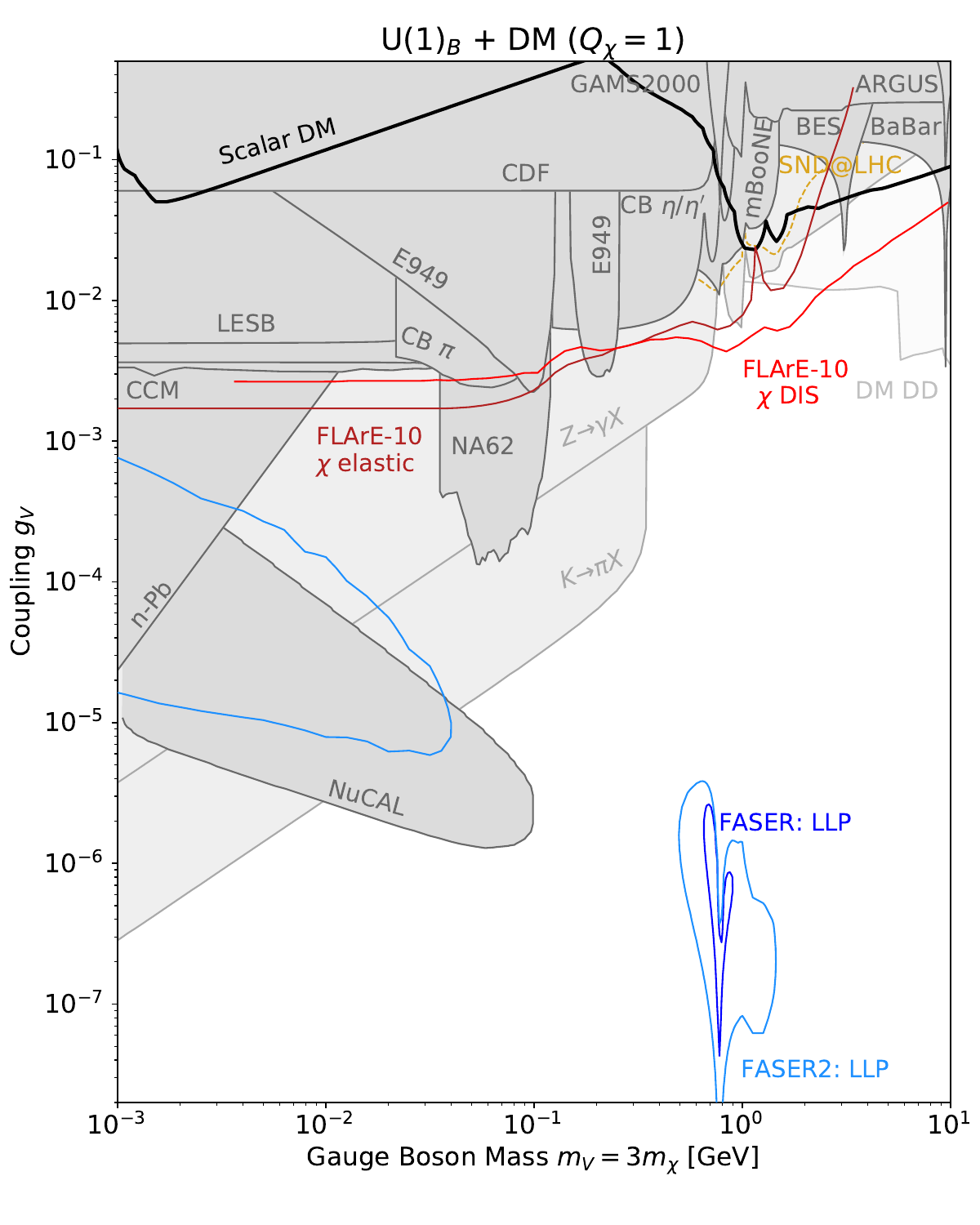}
    \includegraphics[width=0.49\textwidth]{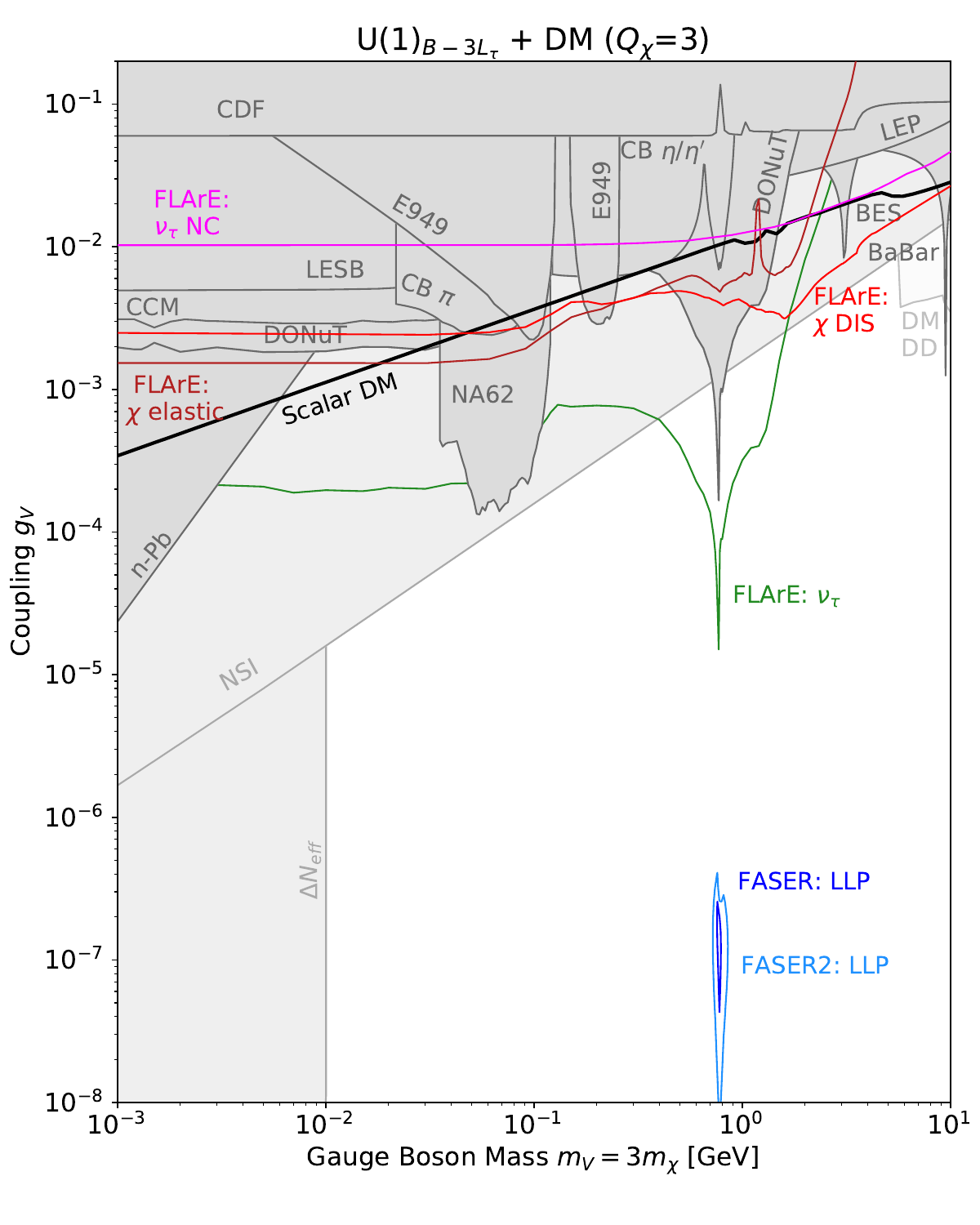}
    \caption{
    Same as \cref{fig:fixedalphaD}, but for only the FLArE-10 detector, complex scalar DM, and fixed charges $Q_\chi = 1$ (left) and $Q_\chi = - Q_\tau = 3$ (right), resulting in a floating $\alpha_\chi$. Additional expected exclusion bounds from probing displaced $V$ decays to SM final states in FASER (FASER2) are shown with dark (light) blue lines. In the right panel, the green contour is the sensitivity contour from probing excess CC scatterings of $\nu_\tau$.
    }
    \label{fig:gequal}
\end{figure*}

In \cref{fig:fixedalphaD}, we also present the expected $90\%$ CL exclusion bounds in searches for DM scatterings off nuclei in the elastic (dark red) and DIS (light red) 
channels for FLArE-10 (solid), FLArE-100 (dash-dotted), and FASER$\nu$2 (dotted). As is clear from the plot, the elastic scattering probe is stronger for light DM and mediator masses below $1~\gev$, which favor interactions with low momentum exchange. For $m_V\gtrsim 1~\gev$, the elastic scattering rate is suppressed by the form factor and the cut on the outgoing proton momentum $p_p\lesssim 1~\gev$. In this higher mass range, the search based on DIS processes provides the best reach. For comparison, we also show the expected reach of the SND@LHC detector~\cite{Boyarsky:2021moj} with the assumptions noted in \cref{sec:detectors}.

For the U(1)$_B$ model with fixed $\alpha_\chi = 0.01$ shown in the upper left panel of \cref{fig:fixedalphaD}, we expect that future searches at the FPF will cover almost the entire remaining allowed region in the parameter space above the Majorana and complex scalar relic target lines, in which DM is not thermally overproduced in the early Universe. This corresponds to vector boson masses between $1$ and $3~\gev$. For the simple complex scalar DM model, additional stringent bounds for $m_\chi\gtrsim 200~\mev$ can arise from past DM DD searches, which are indicated in the plots by the very light gray shaded regions and cover the region within the sensitivity of FLArE and FASER$\nu$2. However, these limits can be evaded in the inelastic scalar DM case and are not relevant for Majorana DM. For lower masses, $(\textrm{a few})~\mev \lesssim m_V\lesssim 1~\gev$, the expected FLArE and FASER$\nu$2 bounds extend beyond current constraints from the CCM, MiniBooNE, and NA62 experiments. Here, the searches at the FPF would probe regions in the parameter space that are otherwise partially excluded only by anomaly-induced rare $K$ and $Z$ decays. 

Next, we consider the U(1)$_{\btau}$ model with fixed $\alpha_\chi = 0.01$ shown in the upper right panel of \cref{fig:fixedalphaD}. Since the model is free of gauge anomalies, the stringent constraints from rare $Z$ and meson decays present in the U(1)$_B$ model are absent in this case. On the other hand, the additional bounds from neutrino NSI cover much of the model parameter space. Nevertheless, we observe that the FPF detectors can still explore a portion of the currently allowed parameter space, especially in the $\omega$ and $\phi$ resonance regions, $m_V\sim m_\omega, m_\phi$, and the corresponding part of the relic target line for complex scalar DM. In this model, additional sensitivity arises from dark vector boson-mediated scattering of tau neutrinos in the DIS regime; see \cref{sec:BSMneutrinos}. The relevant expected bounds, which are indicated by the light purple lines in the plots, impact parameter regions that are already excluded by past searches. We note that the actual exclusion bound in the DIS channel should be derived using the combined excess signal rates for both DM and BSM neutrino scatterings over the expected SM backgrounds. Instead, in the plot, we have presented the expected bounds for each separately to allow for independent discussion of the impact of different new physics effects.

For larger values of $\alpha_\chi$, the relic target lines shift downwards relative to the FPF sensitivity contours from DM scattering. This is dictated by the different parametric dependence of the annihilation cross section and the number of DM scattering events in the FPF experiments on the coupling constants, $\langle\sigma v\rangle\sim g_V^2\alpha_\chi$ and $N_{\textrm{ev}}\sim g_V^4\alpha_\chi$, respectively. As a result, in the lower panel of \cref{fig:fixedalphaD} obtained for $\alpha_\chi=0.5$, we observe that both FLArE and FASER$\nu$2 will only partially cover the thermal target lines for the U(1)$_B$ model. Instead, in the U(1)$_{\btau}$ case, they will typically probe regions in the parameter space predicting subdominant fractions of thermally-produced $\chi$ DM.

Thus far we have considered scenarios in which the vector boson mediator couples much more strongly to DM than to SM particles, $Q_\chi \gg 1$. In \cref{fig:gequal} we consider the different scenario in which the vector boson mediator couples with comparable strength to complex scalar DM and SM particles, with $Q_\chi$ fixed according to \cref{eq:Qchi}. As can be seen, for both the $U(1)_B$ and $U(1)_{B-3\tau}$ models, FLArE-10 can cover the entire relic target line in a wide vector boson mass range between $1~\mev$ and $10~\gev$. As in the previous scenarios depicted in \cref{fig:fixedalphaD}, significant portions of these regions are already constrained by either anomaly-induced or NSI bounds, as well as by the other past searches indicated in the plots. However, we emphasize that for the case of inelastic scalar DM in the U(1)$_B$ model, to which DD constraints do not apply, FLArE-10 will be able to test an interesting open region of parameter space for vector boson mass of order several GeV that is consistent with the observed DM abundance. 

\cref{fig:gequal} also highlights the rich phenomenology present in scenarios with comparable DM and SM couplings to the vector boson mediator. Along with the scattering searches relevant for DM and BSM neutrino interactions, additional prospects arise at very small couplings from FPF searches for visible decays of the long-lived vector boson mediator; see \cref{sec:proddecay}. In particular, for $m_V$ between several hundred MeV and a GeV and coupling $10^{-8}  \lesssim g_V \lesssim 10^{-5}$, such displaced decays into visible final states, primarily light hadrons, can be detected at both FASER and FASER2. The dominant branching fraction in this case is into three pions, $\pi^0\pi^+\pi^-$, which leads to a striking signature consisting of a photon pair accompanied by two oppositely charged tracks. 

We present the relevant expected $90\%$ CL exclusion bounds on LLP decays for FASER (dark blue) and FASER2 (light blue) in the plots. These correspond to the region of parameter space with $m_V\sim m_\omega$ or $m_\phi$. Here, both the dark vector boson production via proton bremsstrahlung and its decay branching fractions into light hadrons are enhanced. The expected exclusions shown in the plot are bounded from below by the production rate of the dark vector bosons being too low, and from above by the $V$ lifetime being too small for the boson to decay in the detectors. In the U(1)$_B$ model, further sensitivity at FPF experiments can be obtained for $m_V\lesssim 10~\mev$ due to loop-induced dark vector boson decay into an $e^+e^-$ pair. This scenario is, however, already constrained by the past beam-dump search in NuCal and by the anomaly-induced bounds.

Last but not least, in the U(1)$_{\btau}$ model, further constraints arise due to the dominant dark vector boson decays into tau neutrinos. These can generate an excess flux of $\nu_\tau$s over the expected SM production rates, which can be detected via their CC scatterings in the detector, as described in \cref{sec:tausignature}. The corresponding expected sensitivity is indicated by the green contour in the right panel of \cref{fig:gequal}. For $m_V\lesssim 2~\gev$, this sensitivity is greater than from the DM and BSM neutrino searches. In particular, it allows one to constrain the currently allowed region of the parameter space of the model close to the $\omega$- and $\phi$-resonance regions. In this case, the increased flux of $\nu_\tau$s could also further contribute to the aforementioned NC DIS signal rate due to BSM tau neutrino interactions. To isolate the impact of various new physics effects, we do not take this into account when presenting relevant expected bounds, which should thus be considered conservative. We stress that the dominant expected bound in the corresponding region of the parameter space is, in any case, due to excess CC $\nu_\tau$ scatterings.

\section{Conclusions}
\label{sec:conclusions}

While the ability of the LHC to search for TeV-scale DM is well known, recently proposed dedicated experiments at high rapidity can significantly enhance the potential of the LHC to probe light DM. Beyond the minimal portal extensions of the SM that allow for light DM and an associated mediator, new gauge groups represent a well-motivated class of possible dark sector models. In this paper, we have explored the use of the FPF to study such U(1) theories leading to hadrophilic dark sectors. In particular, these remain beyond the reach of experiments focusing on BSM electron couplings, while they can more straightforwardly be studied at the LHC.

The suite of FPF experiments provides a comprehensive set of tests of these theories in different regions of parameter space. DM produced in $pp$ collisions can scatter in the FASER$\nu$2 and FLArE detectors through the new light vector boson; we have considered both elastic and deep inelastic scattering. Furthermore, if the mediator has a significant decay branching ratio to SM states, the FASER2 LLP detector can search for the visible decay products. In fact, already at Run 3, the FASER detector will begin to test hadrophilic U(1) theories at couplings substantially lower than existing bounds. These hadrophilic models therefore motivate near-term searches at FASER for new LLP signatures
\begin{equation}
V \to \pi^0 \gamma, \, \pi^+ \pi^- \pi^0, \, K^+ K^-, \, K_S K_L  \ ,
\end{equation}
which are not motivated by dark photon models for FASER in Run 3. Finally, if neutrinos are also charged under the new gauge symmetry, additional signatures are possible in the scattering detectors. We have demonstrated that with a symmetry under which tau neutrinos are charged, the $\nu_\tau$ flux and NC cross section are both enhanced, leading to potential deviations in $\nu_\tau$ CC and NC scattering rates.

These results for U(1)$_B$ and U(1)$_{\btau}$ models should be considered as illustrative of the complementarity of forward LHC experiments in searching for light dark sectors, particularly between LLP and scattering detectors. In both of these theories, the FPF can test broad regions in the coupling-gauge boson mass plane, including significant expanses over which the observed DM relic density could be obtained through standard thermal freezeout. For our benchmark scenario with scalar DM, $m_V = 3 m_\chi$ and low values of the dark charge $Q_\chi$ given by \cref{eq:Qchi}, FPF searches can probe well below the thermal relic target lines in each model for nearly all gauge boson masses between 1 MeV and 10 GeV. Throughout our results, the strongest searches tend to be those based on DM elastic scattering and DIS, with distinct additional reach possible from LLP searches when the mediator can decay to SM final states. For the $\btau$ model, searches for an increased $\nu_\tau$ flux would also test new space. 

The models that we have studied face strong indirect constraints, notably from rare invisible decays and neutrino oscillations, but we emphasize that FPF searches can test couplings that are smaller than these formidable existing bounds. In addition, in the GeV mass range, these searches provide constraints that are complementary to those from spin-independent DD, the latter only being applicable in the case of elastic scalar DM.

Though we have chosen to focus on two possible gauge groups with a handful of coupling and mass assumptions, the general interplay between the DM scattering, LLP and neutrino searches is likely to persist for other theories and parameter choices. To determine the gain provided by the FPF in a particular theory, the reach of these searches must be compared against those from other bounds. As we have seen, U(1) theories that are not anomaly-free typically face rare meson decay constraints, while those with nonzero lepton charges can encounter NSI bounds. For models with couplings to 1st and 2nd generation leptons, additional limits from beam dump and neutrino experiments would likely need to be considered as well.

Forward LHC detectors offer a distinct perspective on light hidden sectors, allowing for searches for light DM and its associated mediators in an otherwise inaccessible kinematic regime. The results here underscore the utility of different types of forward detectors, as could be provided at the FPF. The multi-pronged approach to uncovering physics beyond the SM that is enabled by such a facility, along with other uses such as measurements of SM neutrino interactions and tests of QCD, bolsters the physics case for the FPF.

\acknowledgements

We thank Daniele Alves, Asher Berlin, Patrick deNiverville, Peter Reimitz, and Tyler Thornton for useful discussions and correspondence.  We are also grateful to the authors and maintainers of many open-source software packages, including 
\texttt{FeynCalc}~\cite{Shtabovenko:2020gxv}, 
\texttt{GENIE}~\cite{Andreopoulos:2009rq,Andreopoulos:2015wxa},
\texttt{LHAPDF}~\cite{Buckley:2014ana},
\texttt{scikit-hep}~\cite{Rodrigues:2019nct}. The work of B.B.~is supported by the U.S. Department of Energy (DOE) under grant No.~DE–SC0007914. The work of J.L.F.~is supported in part by U.S.~National Science Foundation (NSF) Grant Nos.~PHY-1915005 and PHY-2111427 and by Simons Investigator Award \#376204. M.F. is supported in part by NSF Grant Nos.~PHY-1915005 and by NSF Graduate Research Fellowship Award No.~DGE-1839285. A.I.~and R.M.A.~are supported in part by the DOE under Grant No.~DE-SC0016013. R.M.A.~is supported in part by the Dr.~Swamy Memorial Scholarship. The work of F.K.~is supported by the DOE under Grant No.~DE-AC02-76SF00515 and by the Deutsche Forschungsgemeinschaft under Germany’s Excellence Strategy -- EXC 2121 Quantum Universe -- 390833306. S.T.~is supported by the grant ``AstroCeNT: Particle Astrophysics Science and Technology Centre'' carried out within the International Research Agendas programme of the Foundation for Polish Science financed by the European Union under the European Regional Development Fund. S.T.~is supported in part by the Polish Ministry of Science and Higher Education through its scholarship for young and outstanding scientists (decision no 1190/E-78/STYP/14/2019). S.T.~is also supported in part from the European Union’s Horizon 2020 research and innovation programme under grant agreement No.~952480 (DarkWave project).


\bibliography{references}

\end{document}